# Optimal Capacity of a Battery Energy Storage System based on Solar Variability Index to Smooth out Power Fluctuations in PV-Diesel Microgrids

Julius Susanto, *Member*, *IEEE*, and Farhad Shahnia, *Senior Member*, *IEEE*

*Abstract*—Battery energy storage systems can be integrated with photovoltaic (PV)-diesel microgrids, as an enabling technology to increase the penetration of PV systems and aid microgrid stability by smoothing out the power fluctuations of the PV systems. This paper has focused on this topic and aims at deriving correlations between the optimal capacity of the smoothing batteries and variabilities in daily solar irradiance. To this end, two most commonly used techniques of moving average and ramp rate control are employed on a real solar irradiance dataset with a 1-minute resolution for a full calendar year across 11 sites in Australia. The paper then presents the developed empirical model, based on linear regressions, to estimate the optimal capacity of the batteries without requiring the use of detailed simulation studies, which are useful for practitioners at the early stages of a project's feasibility evaluation. The performance of the developed technique is validated by numerical simulation studies in MATLAB®. The study demonstrates that the empirical model provided reasonably accurate estimates when using the moving average smoothing technique, but had limited accuracy under the ramp rate control technique.

NOMENCLATURE

*A. Abbreviations*

| | |
|---|---|
| CDF | Cumulative distribution function |
| DoD | Depth of discharge |
| GHI | Global horizontal irradiance |
| MA | Moving average |
| PONE | Probability of non-exceedance |
| PV | Photovoltaic |
| RR | Ramp rate |
| SB | Smoothing battery |
| SBOC | Smoothing battery's optimal capacity |
| SIVI | Solar irradiance variability index |
| SoC | State of charge |

*B. Parameters and Variables*

| | |
|---|---|
| $E_{SB}^{nom}$ | SB's nominal capacity |
| $(E_{SB}^{nom})^{pred}$ | SB's nominal capacity defined by linear regression |
| $k_{RRL}$ | RR limit |
| $N_w$ | MA window size |
| $P_{MA}$ | Smoothed output using MA technique |
| $P_{PV}$ | Raw (unsmoothed) output from PV system |

The authors are with the Discipline of Engineering and Energy, Murdoch University, Perth, Australia.
(*Corresponding Author: susanto@ieee.org)

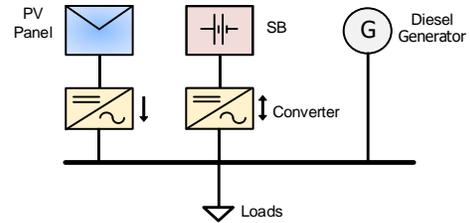

Fig. 1. A typical PV-diesel microgrid with a power smoothing battery (SB).

| | |
|---|---|
| $P_{RR}$ | Smoothed output using RR control technique |
| $P_{SB}^{MA}$ | SB's net output under the MA control technique |
| $P_{SB}^{RR}$ | SB's net output under the RR control technique |
| $SBOC^{ap}$ | Approximate SBOC |
| $\Delta P_{max}$ | Maximum allowable change in PV output in a time step (based on RR control limit) |
| $\Delta T$ | Averaging interval for SIVI calculation |

## I. INTRODUCTION

THERE are thousands of standalone remote area power systems and microgrids around the world that supply remote and island communities with no access to the grid. Most of these systems rely solely on diesel generators [1]. Yet diesel generators are expensive and in recent years, with the rapid decline in the costs for photovoltaic (PV) systems and balance of plant, PV systems have become significantly more competitive than diesel on a levelized cost of energy basis [2]. There is, therefore, growing interest in hybridizing diesel-based microgrids with PV systems, and maximizing the penetration of the PV component (see Fig. 1).

However, there are technical limits to the penetration of PV systems into a diesel microgrid, which are associated with factors such as frequency stability and under-loading of the diesel generators [3]. One option to increase the penetration of PV systems is to use enabling technologies such as battery energy storage systems to smooth out the fluctuations in the output power of PV systems [4-6], thereby limiting frequency deviations and reducing the wear and tear on the diesel generators from excessive ramping [7]. The use of energy storage systems is commonly used for integrating renewable energy resources into microgrids [8-9].

Traditionally, the sizing of battery energy storage systems for smoothing the power fluctuations of PV systems has been achieved by chronological/time-sequential simulation studies, where a smoothing algorithm is applied to the PV system's output and the energy exchange between the battery and the power system is calculated at each time step. Similar methods are also used in the smoothing of wind power fluctuations [20-25]. For the smoothing of PV fluctuations, the application of high-resolution solar irradiance data (i.e., less than 1-min) is necessary to adequately reflect the extents of short-term variability in solar irradiance [26]. However, several previous studies have used coarse 1-hour resolution data, which would introduce significant temporal smoothing from the solar irradiance data alone [13, 16]. Moreover, data for at least a year should be used for sizing a smoothing battery (SB) to capture seasonal variations in solar irradiance variability [11, 13]. Some studies such as [10, 17] have only considered data for one day, while [18] has examined only 1-hour of data. Finally, while some studies such as [13-14, 18] make use of fully specified battery capacity models, many researchers have not considered the battery charge/discharge characteristics and they have sized the battery capacity based on either the peak energy exchange [15-16] or the net energy exchange [17]. Furthermore, in some studies, there is also no consideration for key parameters such as the battery's depth of discharge (DoD) and the initial state of charge (SoC) [15, 17]. It should be noted, however, that many of these previous works are predominantly concerned with the control and performance aspects of different smoothing algorithms, and battery sizing is not their main emphasis. Table I summarizes the key features of the above-mentioned studies (i.e., using low/high resolution and the considered period of solar irradiance data, considering various geographic locations, and employing a detailed battery model).

Against this backdrop, this paper aims to address some of the shortcomings of previous works by developing a comprehensive methodology for the optimal sizing of SBs, employed in PV-diesel microgrids, using one year of high-resolution solar irradiance data across multiple geographically diverse sites, and using a fully specified battery capacity model. Furthermore, this paper investigates the correlations between the SB's optimal capacity (SBOC) and the solar irradiance variability index (SIVI), and proposes an empirically derived estimate for the optimal battery size that can be calculated without simulation studies. Table I also shows the key features of the proposed technique in this paper.

In summary, the main contributions of this work are:
- Proposing to consider the impact of SIVI in determining the SBOC in PV-diesel microgrids, and determining their correlation,
- Developing empirical estimates, based on linear regressions, to estimate the SBOC by only considering the SIVI and without requiring the use of detailed simulation studies, and
- Defining the sensitivity of the SBOC versus the battery and ambient parameters.

The remainder of the paper is organized as follows: Section II discusses how solar irradiance fluctuations throughout a day can be measured and quantified with the SIVI. The considered PV-diesel microgrid with the proposed SB is introduced in Section III. This section also discusses the proposed approaches to determine the SBOC. The relationships between the SBOC and the SIVI are derived in Section IV through numerical analyses over a real dataset from 11 weather stations around Australia. The performance of the proposed technique versus two existing approaches in the literature are presented in Section V. Section VI discusses the sensitivity of the SBOC against the key system design parameters while the practical considerations and limitations of the SBOC determination are briefly discussed in Section VII. Finally, the key findings of the research are summarized and highlighted in Section VIII while two appendices provide details of the employed modeling approach and technical parameters in the studies of this paper.

Table I. The key features in the existing literature on sizing of SBs.

| Ref. | High-resolution data | ≥1 year of solar data | Multiple locations | Battery model |
|---|---|---|---|---|
| [7, 10] | ✓ (20 sec) | ✗ (1 day) | ✗ | ✗ |
| [11, 12] | ✓ (5 sec) | ✓ (1 year) | ✗ | ✗ |
| [13] | ✗ (1 hour) | ✓ (1 year) | ✓ (two) | ✓ (Kinetic) |
| [14] | ✗ (1 hour) | ✓ (1 year) | ✓ (two) | ✓ (Kinetic) |
| [15] | ✓ (1 min) | ✗ (1 day) | ✗ | ✗ |
| [16] | ✗ (1 hour) | ✗ (1 day) | ✗ | ✗ |
| [17] | ✓ (30 sec) | ✗ (1 day) | ✗ | ✗ |
| [18] | ✓ (≤1 min) | ✗ (1 hour) | ✗ | ✓ (Internal resistance) |
| This paper | ✓ (1 min) | ✓ (1 year) | ✓ (11) | ✓ (Kinetic) |

Table II. Considered study sites in this study.

| Site | Name | State | Longitude (°) | Latitude (°) |
|---|---|---|---|---|
| 1 | Adelaide | South Australia | -34.9285 | 138.6007 |
| 2 | Alice Springs | Northern Territory | -23.6980 | 133.8807 |
| 3 | Rockhampton | Queensland | -23.3791 | 150.5100 |
| 4 | Cape Grim | Tasmania | -40.6833 | 144.6833 |
| 5 | Kalgoorlie | Western Australia | -30.7490 | 121.4660 |
| 6 | Darwin | Northern Territory | -12.4634 | 130.8456 |
| 7 | Broome | Western Australia | -17.9614 | 122.2359 |
| 8 | Learmonth | Western Australia | -22.2312 | 114.0888 |
| 9 | Geraldton | Western Australia | -28.7774 | 114.6150 |
| 10 | Wagga | New South Wales | -35.1082 | 147.3598 |
| 11 | Townsville | Queensland | -19.2590 | 146.8169 |

## II. SOLAR IRRADIANCE VARIABILITY

One of the key features of this paper is the explicit consideration of the SIVI for sizing an SB. This is because various geographical locations are subject to different levels of SIVI over a calendar year, and consequently, the sites with low SIVI do not require as much smoothing as the sites with higher SIVI.

The solar irradiance dataset, used in this study, is based on the 2017 measurement data of the global horizontal irradiance (GHI), with a 1-minute resolution, captured from the Australian Government's Bureau of Meteorology [27]. This data was retrieved for 11 weather station locations around Australia, as listed in Table II. The selected areas vary from urban (site-1, 6) to rural (site-4, 8) locations, coastal (site-7, 8, 9) to inland areas (site-2, 5), and hot (site-3, 11) to cold (site-4, 10) regions.

Fluctuations of the solar irradiance throughout a full day, at a specific location, can be quantified as a single number using the SIVI in the form of [28]

$$SIVI = \frac{\sum_{t=1}^{T-1} \sqrt{[GHI(t+1) - GHI(t)]^2 + \Delta T^2}}{\sum_{t=1}^{T-1} \sqrt{[GHI_{\text{clear-sky}}(t+1) - GHI_{\text{clear-sky}}(t)]^2 + \Delta T^2}} \quad (1)$$

Table III. SIVIs for various PONE levels for site-1 to 4 of Table II in 2017.

| PONE | site-1 | site-2 | site-3 | site-4 |
|---|---|---|---|---|
| P50 | 4.8 | 2.0 | 8.4 | 8.2 |
| P75 | 7.2 | 8.4 | 12.5 | 12.1 |
| P90 | 11.1 | 10.2 | 15.3 | 15.4 |
| P95 | 12.4 | 12.7 | 17.8 | 18.5 |
| P99 | 22.1 | 21.3 | 22.5 | 23.4 |
| P100 | 26.3 | 25.0 | 25.0 | 32.9 |

where $GHI$ is the measured global horizontal irradiance (W/m$^2$), $GHI_\text{clear-sky}$ is the clear-sky irradiance and $\Delta T$ is the averaging interval while $T$ is the number of consecutive measurements (e.g., given a minute-averaged time series, $\Delta T = 1$ and $T = 1440$ for a 24-hour period).

The SIVI provides a useful measure for classifying different days based on their solar intermittency relative to the expected clear-sky irradiance profile. An SIVI of close to unity represents the ideal clear-sky day (refer to Fig. 2a, as an example, captured from site-1 on 25/01/2017), while a high SIVI is more representative of a mixed-sky day (refer to Fig. 2b, captured from the same site on 06/11/2017). Note that an overcast day may have a low SIVI despite having uniformly low irradiance throughout the day (as an example, see Fig. 2d, captured from the same site on 16/07/2017).

The SIVI for a single location can be calculated for each day in the available dataset with a minimum of one year to cover seasonal variations and combined to form empirical non-parametric cumulative distribution functions (CDFs), instead of the true CDF. The CDFs provide insights into a site's SIVI over time.

Fig. 3 illustrates the empirical CDFs for site-1 to 4 of Table II in 2017. As seen in Fig. 3b, Site-2 shows SIVIs that are close to unity for over 40% of the year, indicating a site that has consistent clear sky days. On the other hand, as seen in Fig. 3c, Site-3 exhibits a much higher spread in the SIVI, suggesting a site with a higher prevalence of mixed-sky days.

The SIVI data in the CDFs can also be represented as a probability of non-exceedance (PONE) values, indicating the probability that an SIVI is not exceeded. For example, Table III provides the SIVI values under various PONE limits for the 4 sites of Fig. 3. As seen from this table, at site-2, a P90 value of 10.2 denotes that the SIVI does not exceed 10.2 for 90% of the year. The PONE limits will be used in the remainder of the paper to select an appropriate smoothing level for the SBs at various locations.

### III. SIMULATION MODEL AND OPTIMIZATION

The system topology, considered in this article, is an ac-coupled PV-diesel microgrid supplying ac loads, as shown in Fig. 1. The SB's functionality in such systems is to smooth out the fluctuations in the output power of the PV systems in order to limit the ramping requirements and stresses on the diesel generator(s). This is crucial for increasing the PV penetration in the microgrid without causing frequency instability and poor frequency regulation.

For the purposes of this study, neither the diesel generator nor the loads are explicitly modelled as the study specifically looks at the SB's performance in producing a less volatile output from the solar PV system. It is assumed from the design of the microgrid that the diesel generator and PV system have been adequately sized to supply the load at all times without under- or over-loading the diesel generator. Based on the diesel generator specifications, the maximum ramp-rate capability of the generator defines the amount of smoothing required by the SB [29].

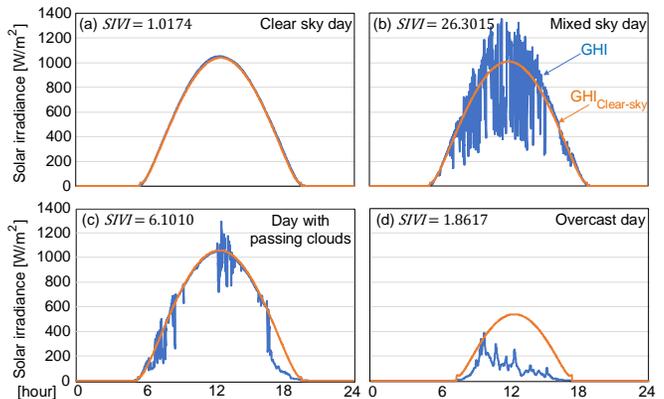

Fig. 2. Solar irradiance and the corresponding $SIVI$ for 4 sample days at site-1: (a) a clear sky day, (b) a day with frequent cloud movements, (c) a clear day with a few passing clouds throughout the day and an overcast period in the afternoon, (d) an overcast day.

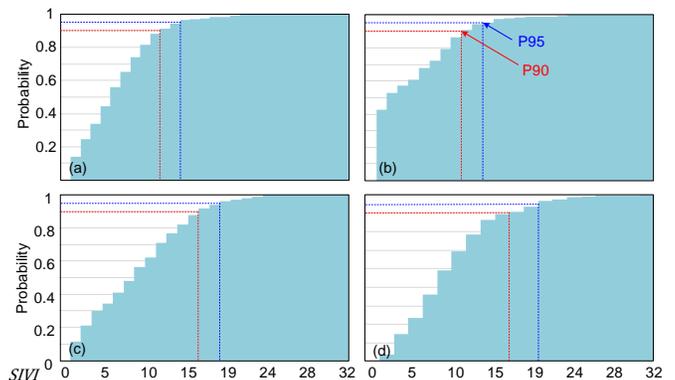

Fig. 3. Empirical CDF for $SIVI$ in 2017 at (a) site-1, (b) site-2, (c) site-3, (d) site-4.

#### A. Power Smoothing Algorithms

Several methods are existing in the literature about power smoothing algorithms via batteries [11-12, 30-31], among which two commonly used techniques are considered here, as discussed below:

#### 1. Moving average (MA) technique

The lagging MA-based smoothing technique takes the arithmetic mean of the PV system's output power for the previous $N_w$ time steps from the current time step of $t$ as [30-31]

$$P_\text{MA}(t) = \frac{\sum_{i=0}^{N_\text{w}-1} P_\text{PV}(t-i)}{N_\text{w}} \qquad (2)$$

where $P_\text{MA}(t)$ is the smoothed PV power output at time step $t$ [W], $P_\text{PV}(t)$ is the raw PV power output at that time step [W] and $N_\text{w}$ is the window of the MA, measured as the integer number of time steps.

*2. Ramp rate control (RR) technique*

The RR-based smoothing technique limits the change in the output power between 1-minute time steps to a maximum value. In this study, the maximum allowable change in the output power is expressed as a proportion relative to the nominal output power of the PV system, in the form of

$$\Delta P_{\max} = k_{\text{RRL}} \times P_{\text{nom}} \quad (3)$$

where $\Delta P_{\max}$ is the maximum allowed change in the PV power output [W], $k_{\text{RRL}}$ is the ramp rate limit [%] and $P_{\text{nom}}$ is the nominal output of the PV module at standard test conditions in watt peak [Wp]. The target output power by this technique ($P_{\text{RR}}(t)$) is then

$$P_{\text{RR}}(t) \quad (4)$$
$$= \begin{cases} P_{\text{PV}}(t-1) - \Delta P_{\max} & \text{if} \quad |\Delta P_{\text{PV}}| > \Delta P_{\max} \ \& \ \Delta P_{\text{PV}} < 0 \\ P_{\text{PV}}(t) & \text{if} \quad |\Delta P_{\text{PV}}| \leq \Delta P_{\max} \\ P_{\text{PV}}(t-1) + \Delta P_{\max} & \text{if} \quad |\Delta P_{\text{PV}}| > \Delta P_{\max} \ \& \ \Delta P_{\text{PV}} > 0 \end{cases}$$

where $\Delta P_{\text{PV}} = P_{\text{PV}}(t) - P_{\text{PV}}(t-1)$ is the change in the output power of the PV system between time steps $t$ and $t-1$.

At each time step, the SB's net output power ($P_{\text{SB}}^{\text{MA}}(t)$) is the difference between the SB's target output power and the PV's output power at that time step, i.e.,

$$P_{\text{SB}}^{\text{MA}}(t) = P_{\text{MA}}(t) - P_{\text{PV}}(t) \quad (5a)$$

$$P_{\text{SB}}^{\text{RR}}(t) = P_{\text{RR}}(t) - P_{\text{PV}}(t) \quad (5b)$$

for the MA and RR-based techniques, respectively. By convention, a positive power (i.e., $P_{\text{SB}}(t) > 0$) denotes that the SB is discharging, while a negative power denotes the SB's charging.

*B. Optimizing SB's Capacity*

The SBOC can be determined within an optimization problem which determines the SB's best (smallest) nominal energy storage capacity ($E_{\text{SB}}^{\text{nom}}$) as

$$SBOC = \min E_{\text{SB}}^{\text{nom}} \quad (6)$$

which is subject to $SoC_{\min} < SoC(t) \leq SoC_{\max}$ while $SoC_{\min}$ and $SoC_{\max}$ respectively denote the SB's minimum and maximum allowed SoC. $E_{\text{SB}}^{\text{nom}}$ is the nominal capacity of the SB (in kWh), i.e. the maximum amount of energy that the SB can store fully charged. Note that $E_{\text{SB}}^{\text{nom}}$ is simply a decision variable to be optimized; i.e., it is an assumed input value (not calculated) for the optimization problem.

One approach to solve (6) is to iteratively run chronological simulations in order to find the smallest $E_{\text{SB}}^{\text{nom}}$ that sastisfies the SoC constraints. This optimization approach is hereafter referred to as the 'chronological simulation method', as dicussed in Appendix-A.

*C. Proposed Approximate Method based on SIVI*

Although the chronological simulation method is more accurate, it is computationally intensive and not necessarily amenable to practical scenarios, especially when conducting preliminary level project feasibility and screening studies. As such, an approximate method is proposed in this paper that has an order-of-magnitude accuracy but leads to very fast results.

This technique is derived by applying the chronological simulation method for each day (refer to Appendix-A) and across all sites in Table II, and then performing a linear regression on the combined results with the SIVI as the dependent variable. A statistical error term (i.e., the standard error) is also added to the regression to capture the upper envelope of the results. The approximated SBOC based on the linear regression (denoted by $SBOC^{\text{ap}}$) is calculated as

$$SBOC^{\text{ap}} = \alpha \times SIVI + \beta + \sigma \quad (7)$$

where $\alpha$ and $\beta$ are the linear regression coefficients and $\sigma$ is the standard error of the estimate, which is calculated from

$$\sigma = \sqrt{\frac{\sum (E_{\text{SB}}^{\text{nom}} - (E_{\text{SB}}^{\text{nom}})^{\text{pred}})^2}{N_s}} \quad (8)$$

where $E_{\text{SB}}^{\max}$ is the SBOC for a single day while $(E_{\text{SB}}^{\max})^{\text{pred}} = \alpha \times VI + \beta$ is the SB's capacity, predicted by the linear regression, and $N_s$ is the total number of samples.

The SIVI can be estimated based on the site location or nearby sites with complete measurement data. If daily SIVI data is available, then a smoothing level (e.g. P90 or P95) can be selected as described in Section II. For example, if the P90 SIVI is 22 and given the coefficients of $\alpha = 0.0046$, $\beta = 0.0567$ and $\sigma = 0.0315$ (applicable for the MA technique with a 10-min window), the approximate SBOC calculated from (7) will be 0.1894 kWh/kWp (i.e., a SB with $E_{\text{SB}}^{\text{nom}} = 0.1894$ kWh is needed for every one kWp of the PV system).

IV. PERFORMANCE EVALUATION

Let us consider the microgrid of Fig. 1 with a standard crystalline silicon PV array connected via a grid-tied inverter. Also, a deep-cycle valve regulated lead-acid SB is considered in the system. The modeling of the PV system and SB are discussed in Appendix-B while the employed model parameters for the PV system are provided in Table C1 in Appendix-C. The SB's assumed discharge characteristics are based on an Olympic Batteries DC2-500 type battery [39], as listed in Table C2 in Appendix-C. The battery's model is assumed to the Kinetic model and same as the models employed in similar works such as [13] and [14]. The bi-directional converter for the SB is considered to have an efficiency of 94%, based on data from a Tier-1 manufacturer [40]. For the base case optimization studies, the SB's DoD is assumed as 70% and the SB is configured at the beginning of each day with an initial SoC of 80%. It is to be noted that, even though considering different battery and converter characteristics will modify the output numerical results of the study, it will not impact the successful performance of the proposed approximate method for determining the SBOC based on the SIVI. The ambient temperature data is retrieved from the Bureau of Meteorology; however, only historical daytime maximum temperatures for each day were accessible in the public dataset. Therefore, the ambient temperature is assumed to be constant throughout the day and set as the historical maximum tem-perature in the studies. This is a conservative assumption and leads to a higher temperature derating of the PV system's output power. It is noteworthy that, when the study was repeated considering hourly ambient temperature data (based on straight line intra-hour interpolation) instead of a constant ambient temperature for the whole day, the determined SBOC did have a marginal difference. This is mainly because the SBOC is primarily driven by fluctuations in the output power of the PV systems over a typical timescale of 5 to 20 minutes and relative changes in the average ambient temperature over such timescales are not very large.

*A. Single Day Study Results*

The simulation model is run for a 24-hour period for site-1 on 5/2/2017 to illustrate the effects of the MA and RR-based techniques on the SB's output power and the final smoothed output from the combined PV-battery system. The considered day is a mixed-sky day with a moderately high *SIVI* of 20.91. For the single day simulations, an unoptimized SB capacity of 1 kWh is applied in order to facilitate comparisons of the SB's performance between simulation results. Note that selection of an SB of 1 kWh is arbitrarily and is only for demonstrating an example of the time-series plots of the smoothing methods and the effect of different parameters on both smoothing performance and battery usage. The results of the study are demonstrated in Fig. 4 and 5, respectively for the MA and RR-based techniques. For each technique, the study was conducted with two different smoothing parameters (i.e. MA window and RR limit) to demonstrate the impact of more onerous smoothing requirements. As can be seen from these figures, a larger MA window (20 vs 5 min) or a tighter RR limit (1 vs 10%) produces a smoother output, but a larger proportion of the battery capacity is used as a result.

*B. SB Optimization Results*

The chronological simulation method of Appendix-A has been applied for the 11 sites of Table II, for each day over the full calendar year of 2017. Fig. 6a and b illustrate the results for 4 sites when respectively the MA and RR techniques are employed. The study attempts to determine the minimum SBOC, required to meet the selected smoothing criteria, for a given day and site. For each site, the results are presented as a scatter plot where each point represents the SB's optimized capacity for a single day (on the y-axis) and the corresponding SIVI for that day (on the x-axis). An empirical CDF of the optimized SB capacities over the full year is also shown to illustrate how the optimization results are distributed and aid in the SBOC's selection.

The optimization results are expressed in kWh of the SB's capacity per kWp of the PV systems, and can be scaled linearly to any desired PV capacity. As such, if the optimization results yield an SBOC of 0.4 kWh per kWp, then a 100 kWp PV system would require an SB with a capacity of 40 kWh.

The optimization results for the MA-based technique in Fig. 6a illustrates a clear positive correlation between the SBOC and the SIVI. This result is according to the general expectation of observing higher levels of SB's charge/discharge during days of high SIVI and vice versa. However, the results for the RR-based technique in Fig. 6b demonstrate a weaker correlation between the SBOC and the SIVI. Indeed, a comparison of the Pearson correlation coefficients between the SBOC using the MA and RR techniques with the observed SIVI in Table IV indicates that, across all studied sites, the RR-based technique is less strongly correlated than the MA-based technique (e.g., the average correlation coefficient across all sites is 0.7862 when using the MA technique versus 0.5556 when the RR technique).

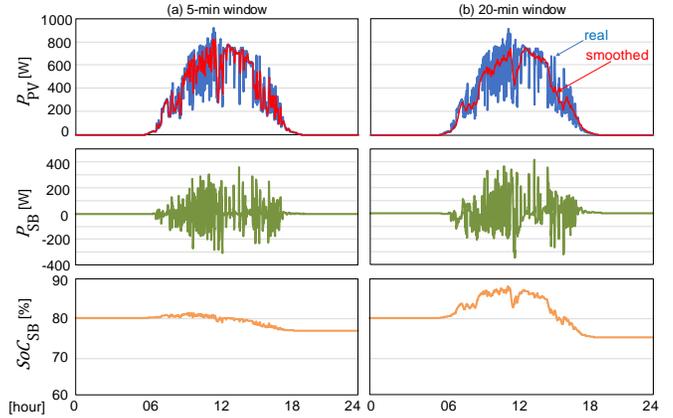

Fig. 4. Performance of the microgrid in Fig. 1 when using an SB, defined under the MA technique for a single day (15/02/2017 at site-1) with a window of (a) 5, (b) 20 minutes.

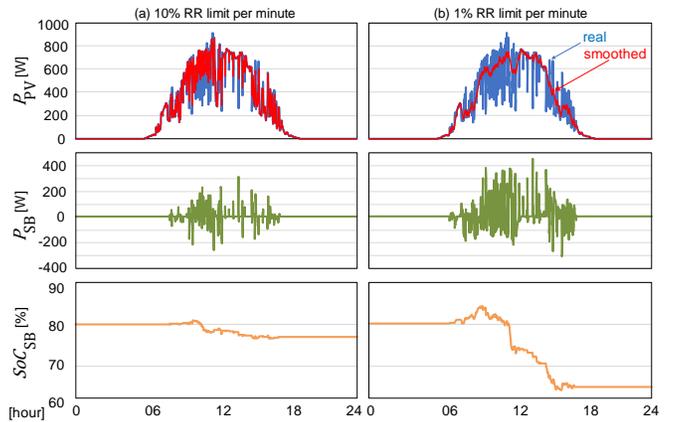

Fig. 5. Performance of the microgrid in Fig. 1 when using an SB, defined under the MA technique for a single day (15/02/2017 at site-1) assuming a window of (a) 10%, (b) 1%.

Table IV. The Pearson correlation coefficient between the determined SBOC and the SIVI for the considered sites in Table I.

| Site | MA (10 min) | RR limit (5%) |
|---|---|---|
| 1 | 0.7786 | 0.5633 |
| 2 | 0.7348 | 0.5750 |
| 3 | 0.8577 | 0.5966 |
| 4 | 0.8180 | 0.6981 |
| 5 | 0.7724 | 0.6914 |
| 6 | 0.7755 | 0.5629 |
| 7 | 0.7302 | 0.4431 |
| 8 | 0.8294 | 0.6383 |
| 9 | 0.7521 | 0.4469 |
| 10 | 0.7849 | 0.4334 |
| 11 | 0.8149 | 0.4621 |
| Average of all sites | 0.7862 | 0.5556 |

Comparing the empirical CDFs for the MA and RR techni-

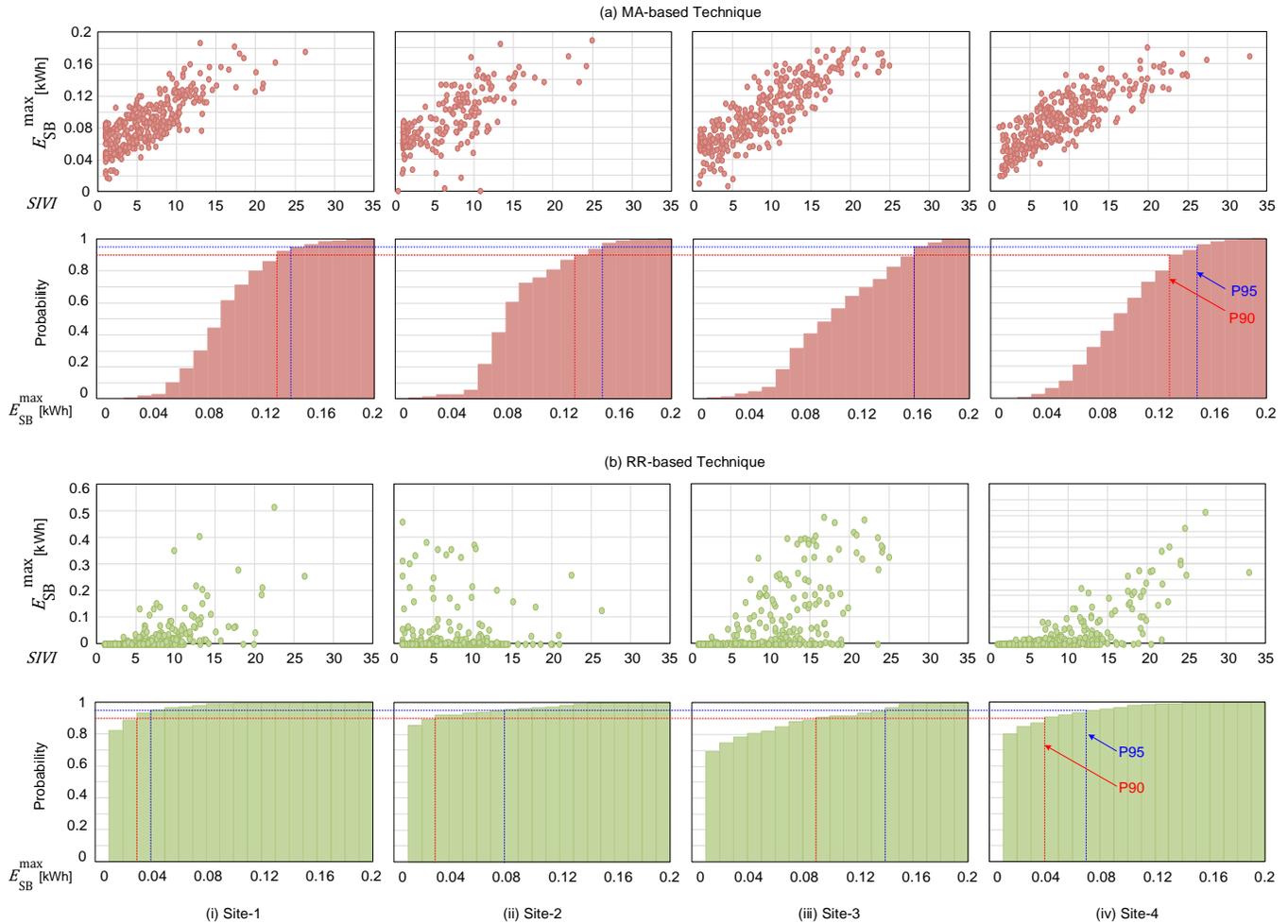

Fig. 6. The SB's optimization results (scatter plot of the SBOC versus *SIVI* and their CDF) using the detailed model with P90 or P95 criterion at site-1 to 4 of Table I while employing (a) the MA-based technique with a 10-min window, (b) the RR-based technique with a limit of 5% of the nominal rating per minute.

ques, respectively given in Fig. 6a and b, indicates that while the SBOC determined by the MA technique appears symmetrically distributed, the RR is positively skewed with the mode centered around zero. This makes intuitive sense as with RR, there are clear-sky days for more than 80% of the studied time (i.e., SIVI close to 1) where no smoothing is required; thus a smaller SBOC is needed. On the other hand, the diurnal nature of the sun results in the MA algorithm to require energy exchange with the SB, even during clear sky days (provided that the MA window is sufficiently large) [31]. This is in line with the finding of [11] which shows increased SB cycling and degradation under the MA technique relative to the RR technique.

To understand why RR is less strongly correlated with SIVI, let us consider the single day simulations at site-5 for two different days of 20/11/2017 and 18/12/2017, illustrated respectively in Fig. 7a and b. In the first day, the SIVI is 10.43 and Fig. 7a shows a mainly clear-sky day with rapidly moving cloud bands occurring in the middle of the day and afternoon. In contrast, the SIVI in the second day is 23.04, and Fig. 7b shows a highly variable mixed-sky day with solar irradiance fluctuations occurring consistently throughout the day. However, the SBOC for the relatively low SIVI at the mainly clear-sky day of Fig. 7a is 0.2837 kWh/kWp, while it is almost 10 times lower (0.029 kWh/kWp) for the high SIVI in the mixed-sky day of Fig. 7b which is in contrary to what is expected.

The difference between the two days is the number of upward and downward ramps. The day with a relatively low SIVI has mainly downward ramps; thus, requiring the SB to discharge for most of the day. On the other hand, the day with a high SIVI has more symmetrical distribution of upward and downward ramps, allowing the SB to both charge and discharge throughout the day. Therefore, for the RR technique, the direction of ramping events is a key factor for sizing the SB. The SIVI provides a crude proxy for the amount of ramping that will be observed since a day with a high SIVI is more likely to exhibit asymmetrical ramping events requiring higher SBOC. However, the outliers in these two figures show that this is not always the case.

Fig. 8 demonstrates the SBOC selected for P95 PONE at the 11 study sites of Table I, using the MA and RR-based techniques. This figure illustrates the results using the chronological simulation method while Table V lists the results from the chronological simulation method, as well as the proposed approximate method, using the linear regression and the observed SIVI, and their difference. The results support the previous observation that MA is strongly correlated with SIVI. The approximate method is inherently conservative because it adds the standard error to the linear regression and the results in Table V show a fairly consistent positive deviation of around 10 to 20%. On the other hand, the large positive and negative deviations found in the results for the RR technique indicate that the approximate method does not accurately predict the SBOC.

Table V. Comparison of the determined SBOC by the detailed and proposed approximate models and their difference, using the P95 desired level for all considered sites in Table I under the 10-min MA and 5% RR techniques.

| Site no. | MA (10-min) | | | RR limit (5%) | | |
|---|---|---|---|---|---|---|
| Model | Detailed [kWh/kWp] | Approximate [kWh/kWp] | Deviation [%] | Detailed [kWh/kWp] | Approximate [kWh/kWp] | Deviation [%] |
| 1 | 0.140 | 0.150 | 7.1 | 0.111 | 0.149 | 34.2 |
| 2 | 0.142 | 0.152 | 7.0 | 0.212 | 0.152 | −28.3 |
| 3 | 0.159 | 0.175 | 10.1 | 0.362 | 0.188 | −48.1 |
| 4 | 0.144 | 0.179 | 24.3 | 0.198 | 0.194 | −2.0 |
| 5 | 0.136 | 0.163 | 19.9 | 0.109 | 0.168 | 54.1 |
| 6 | 0.150 | 0.167 | 11.3 | 0.214 | 0.176 | −17.8 |
| 7 | 0.130 | 0.152 | 16.9 | 0.100 | 0.151 | 51.0 |
| 8 | 0.123 | 0.140 | 13.8 | 0.019 | 0.132 | 594.7 |
| 9 | 0.151 | 0.177 | 17.2 | 0.174 | 0.192 | 10.3 |
| 10 | 0.144 | 0.165 | 14.6 | 0.166 | 0.173 | 4.2 |
| 11 | 0.140 | 0.154 | 10.0 | 0.118 | 0.155 | 31.4 |

Table VI. Coefficients for the suggested approximate method used in the SBOC's calculation.

| Technique | $\alpha$ | $\beta$ | $\sigma$ |
|---|---|---|---|
| MA with 10-min window | 0.0046 | 0.0567 | 0.0315 |
| RR with 5% ramp limit | 0.0074 | -0.0221 | 0.0709 |

Table VII. Comparison of SBOC sizing methods for site-1 using the MA-based technique with a 10-min window.

| | SBOC sizing method | SBOC [kWh/kWp] | Annual Coverage [%] |
|---|---|---|---|
| Existing Methods | Peak energy exchange | 0.131 | 92.8 |
| | Hourly chronological simulation | 0.619 | 100.0 |
| Methods of this paper | 1-min chronological simulation (P95 desired level) | 0.140 | 95.0 |
| | Approximate method (P95 desired level) | 0.150 | 96.5 |

Fig. 9a and b show the linear regression using the MA and RR techniques and with data from all sites that have been used in the study of Fig. 8. The regression coefficients for the MA and RR algorithms, used in (7), are provided in Table VI. It is to be noted that these coefficients are based on model parameters used in this study (i.e., the SB's type, chemistry and DoD, the PV parameters, etc.) and may not be accurate if these parameters are changed.

## V. COMPARATIVE ANALYSIS

To illustrate the superior performance of the proposed technique versus existing similar techniques, the detailed model described in this paper is compared against two SBOC sizing approaches that are commonly found in the literature.

The first method is the peak energy exchange method [15-17]. In this method, the energy exchanged between the SB and the microgrid is integrated over an entire day (allowing for recharging at night), and the maximum value denotes the peak energy exchange for the day.

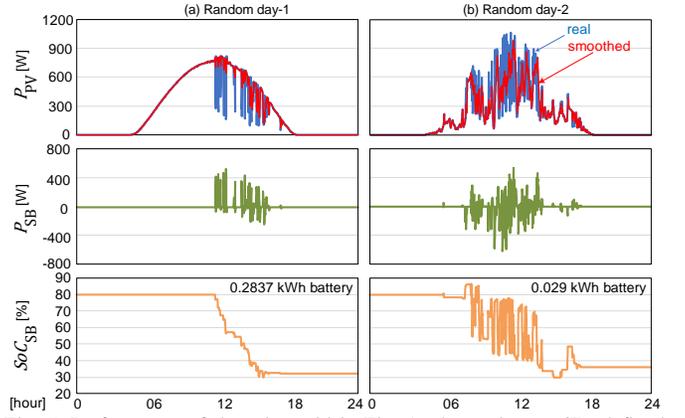

Fig. 7. Performance of the microgrid in Fig. 1 when using an SB, defined under the RR technique with 10% limit at site-5 on two random days of (a) 20/11/2017, (b) 18/12/2017.

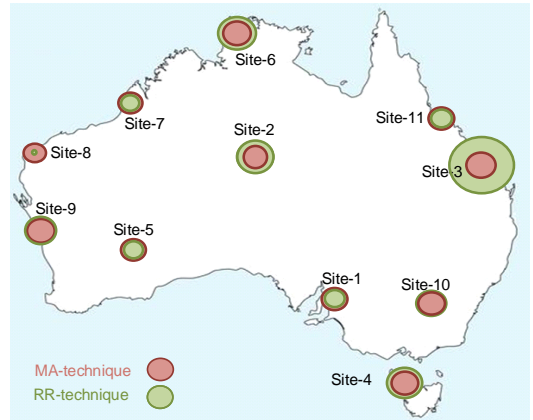

Fig. 8. Comparison of the chronological simulation-based determined SBOC using a 10-min MA and 5% RR techniques for the 11 study sites considered in Table I.

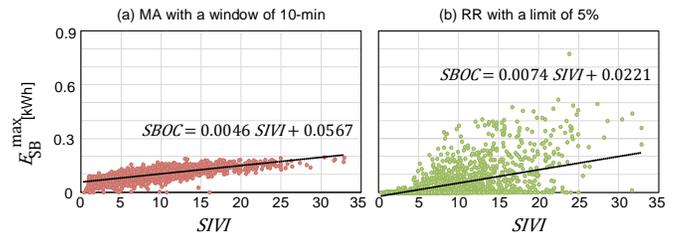

Fig. 9. Linear regression results for all sites using a) MA-based technique with a 10-min window and b) RR-based technique with a limit of 5% of the nominal rating per minute.

This process is repeated for each day of the year, and the maximum value of the year is selected as the SBOC. The second common method is the hourly chronological simulation method [13-14]. This method is structurally similar to the method described in Appendix-A with the key differences being that the sampling resolution is hourly, and the SB is sized to have sufficient capacity for the entire year (i.e., there is no nightly recharging).

The above SB sizing methods are implemented for site-1 using the MA technique with a 10-min window. To facilitate a fair comparison between methods, a maximum allowed DoD of 70% and initial SoC of 80% is applied and nightly recharging is assumed. The results of the comparison are presented in Table VII, showing the SBOC and the implied coverage over the full year, i.e. coverage of 95% means that the SB is sufficiently sized for 95% of the year (347 days). The proposed methods (both detailed and approximate) yield similar SBOC values to the peak energy exchange method, although the SBOC from the peak energy exchange method has lower coverage as this method does not consider the SB's efficiencies or charge/discharge characteristics. Note also that the coverage in the proposed methods is a selected desired value and can be modified based on the SB's design criteria. As can be expected, the hourly chronological simulation method yields a significantly larger SBOC than the other methods, due to the effect of the course 1-hour sampling, resolution filtering out short-term variations in solar irradiance.

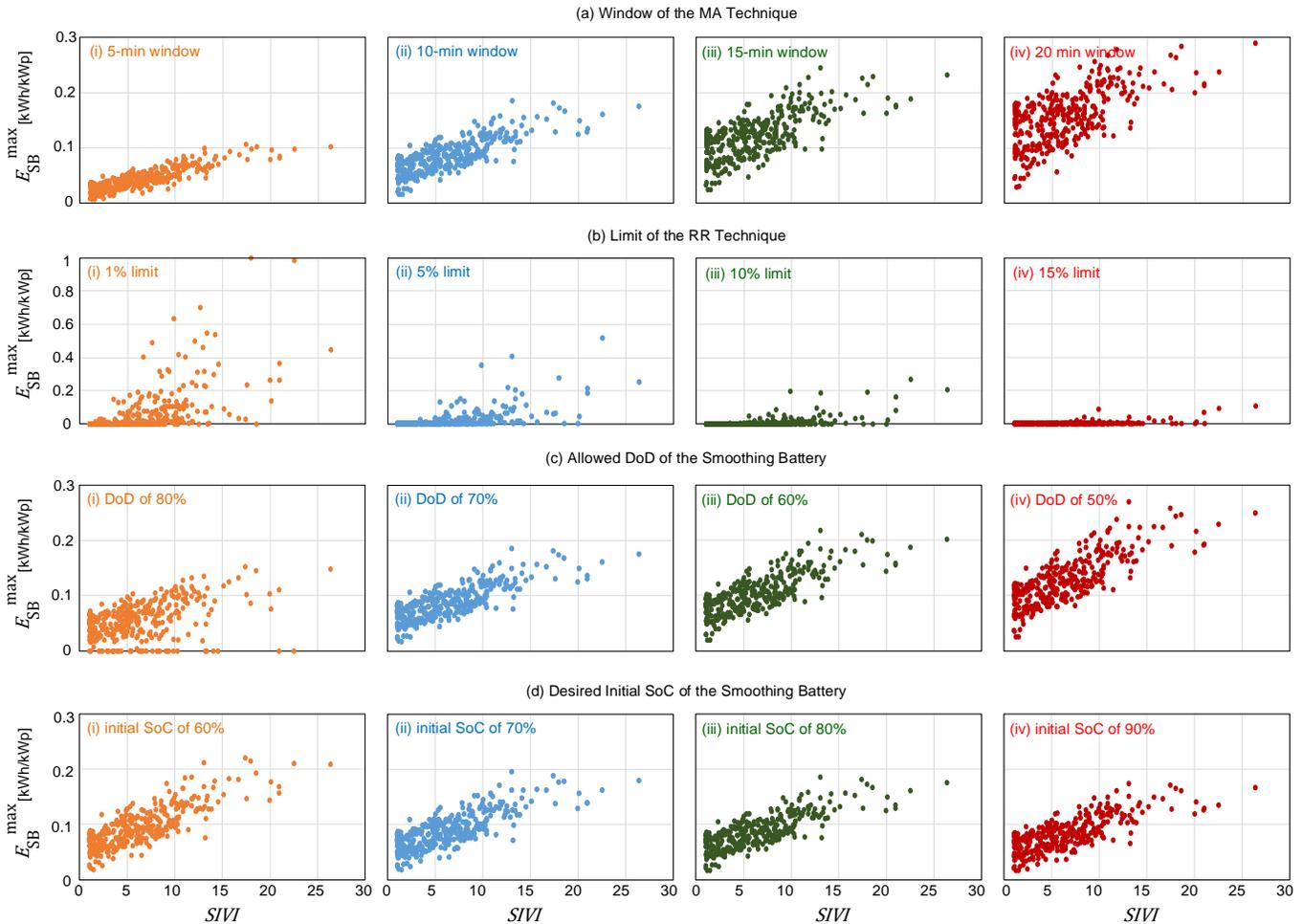

Fig. 10. The sensitivity of the determined SBOC to the (a) window size of the MA technique for a window of 5 to 20 minutes, (b) the limit of the RR technique for a limit of 1 to 15%, (c) a maximum allowed DoD of 80 to 50% for the SB, (d) a desired initial SoC of 60 to 90% for the SB, versus the observed *SIVI*.

The study of Fig. 9a illustrates that the SB sizing under the MA control technique is quite highly correlated to the SIVI of the site. As a result, when determining the SBOC for a specific location, the system planner or designer not need the high resolution irradiance data at the specific site location but can use an estimate of the SIVI or the SIVI measured at a nearby location. This is because, in principle, the linear regression equation of (7) is applicable to any arbitrary site, provided that an estimate of the SIVI is known. This is the key practical use of the proposed approximate method.

## VI. SENSITIVITY ANALYSIS

A sensitivity analysis is carried out to evaluate the variations in the determined SBOC against the key assumptions in the studies; i.e., the window size of the MA technique, the limit of the RR technique, as well as the SB's maximum allowed DoD and the desired initial SoC. The results of these studies are presented in Fig. 10a to d, respectively while in each of the sensitivity analyses, all other key parameters are assumed fixed as per Section IV.A.

### A. Window Size of the MA Technique

As seen in Fig. 10a, as expected, the SBOC increases as the MA window size is expanded from 5 to 20 minutes, with roughly a threefold increase. There is also an increase in the dispersion of the SBOC as the window size is expanded, with $\sigma = 0.0191$ for a 5-minute window and $\sigma = 0.0494$ for a 20-minute window.

## B. Limit of the RR Technique

As shown in Fig. 10b, the SBOC increases as the ramp rate limit decreases from 15 to 1% of the nominal rating per minute. As an example, an average SBOC of 0.0024 kWh/kWp is observed for a 15% ramp rate limit while an average SBOC of 0.0615 kWh/kWp is seen for a 1% ramp rate limit. Similar to the window size of the MA technique, there is an increase in the dispersion of the SBOC as the limit of the RR technique is tightened, with $\sigma = 0.0102$ for a 15% limit and $\sigma = 0.1311$ for a 1% limit.

## C. SB's Maximum Allowed DoD

The maximum allowed DoD may be a material parameter during the design process, as not only does the selection maximum allowed DoD affect battery life, but different battery technologies also have different maximum DoDs. As an example, [32] has reported this parameter varying between 53 and 100% depending on the battery technology. As illustrated by Fig. 10c, the SBOC increases as its maximum allowed DoD decreases from 80 to 50%. At an allowed DoD of 80%, the mean SBOC is 0.0564 kWh/kWp while it increases to 0.1254 kWh/kWp for a DoD of 50%.

## D. SB's Initial SoC

As seen in Fig. 10d, the SBOC increases as the SB's expected initial SoC decreases from 90 to 60%; however, the impacts are relatively small. For example, the mean SBOC is 0.0822 kWh/kWp for an initial SoC of 90% which slightly increases to 0.0949 kWh/kWp for an initial SoC of 60%. As such, through the above sensitivity analyses, it can be seen that the SBOC is largely invariant to the initial SoC and a value of 70-80% is suggested to provide sufficient headroom for the SB's charging and discharging. The sensitivity analyses also show that the maximum allowed DoD has a moderate impact on the SBOC and it is recommended that it is set above 60%, although cycle-life considerations should also be taken into account.

The sensitivity analyses also showed that RR becomes less correlated with SIVI as the ramp rate limit increases, e.g. r = 0.5927 for a ramp rate limit of 1%, and r = 0.43587 for ramp rate limit of 15%. This is because a day with a high SIVI can have many small fluctuations that do not exceed the ramp rate limit, thus the SB does not need to be called into action.

## VII. PRACTICAL CONSIDERATIONS AND LIMITATIONS

This study uses solar irradiance data with a 1-min resolution to capture the short-term temporal SIVI in the output power of the PV system. A question that arises is whether a 1-min resolution adequately covers the range of temporal fluctuations in solar irradiance. For example, [33] reports variations of above 50% in 1-second for a 48 kWp PV system. However, the empirical probability of such variations was found to be 0.000235% (or roughly 2 hours in a year). Ref. [33] also reports that variations of less than 10% in 1-second represent 99.86% of all observations. For 1-min resolutions, variations of less than 10% represent 95.11% of all observations. Based on these findings, it is proposed that 1-min resolution data be considered as the minimum resolution while below 1-min resolutions may be used for higher coverage.

It is worth noting that the Bureau of Meteorology's weather stations are single measurement points and any natural smoothing effects due to spatial diversity [26] are ignored in this study. For small centralized microgrid systems, the absence of spatial smoothing is unlikely to have any significant impact on the results; however, the ignoring the spatial smoothing for larger PV systems or distributed systems will lead to more conservative (less optimal) results.

It is to be noted that differences in weather from one year to another may result in a calculated SBOC that is not truly optimal for future conditions. However, since the SBOC is essentially a capital investment prediction, it needs to be based on historical data. This paper suggests that at least one year of historical data is required to adequately cover seasonal variations. However, it may be more prudent to use multi-year data to capture longer-term (inter-year) variations, such as natural climate cycles, e.g., El Niño and La Niña events. Yet, in practice, this may be contingent on the availability and/or project economics of collecting longer-term data.

## VIII. CONCLUSION

This paper has focused on deriving a correlation between the SBOC and the SIVI, and based on the two most commonly used techniques of MA and RR. The studies were based on a real large solar irradiance dataset with a 1-minute resolution for a full calendar year and across 11 locations. The studies show that the determined SBOC under the MA technique is quite strongly correlated with the daily SIVI, and this observation can be used to develop relatively accurate empirical estimates for the SBOC, which is solely based on the SIVI. A comparison with other SBOC sizing methods indicates that the proposed method is comparable to the peak energy exchange method, but has greater flexibility since the level of coverage is an input variable for the SB's sizing algorithm. Sensitivity analyses also show that the SBOC is not significantly influenced by the SB's initial SoC and maximum allowed DoD. The results of this study also indicate that the determined SBOC by the RR technique is weakly correlated with the SIVI, and as a result, the approximate empirical method using the RR technique-based linear regressions has limited accuracy. As a result, a future research direction can be focused on identifying a more appropriate metric than SIVI to quantify the solar irradiance variabilities when using the RR technique.

## APPENDIX

### A. Chronological Simulation Method

The chronological simulation method used in this paper is a numerical solution to the SBOC optimization problem in (6). The method works by first optimizing the SB for each day and then aggregating the results to select the SBOC.

Note that it is assumed that the SB is recharged (or discharged) each night to maintain a fixed (desired) initial SoC in the morning (e.g., 80%). This assumption is in line with [15] and [16]. This is because the SB capacity can be mini-mized by not having to consider the end-of-day SoC. In fact, the studies of this paper find that it is rare for the end-of-day SoC to be higher than the initial SoC because (a) a preponderance of upward PV output ramps throughout the day, and (b) an SB needs to be recharged at some point (e.g., every night or every week or every month).

*1. Single-day SB Optimization*

The chronological simulation is executed for every simulation day at 1-minute time steps (i.e., 1,440 time steps per 24-hour of the simulation) to capture the variations in the demand and output power of the PV system. At each time step, the computation steps are as below:
a) Retrieving the solar irradiance for the time step from the available dataset,
b) Calculating the expected output power of the PV system from (B1) in Appendix-B.
c) Employing the MA or RR-based smoothing technique and calculating the SB's charging/discharging status and level from (5a) or (5b),
d) Calculating the smoothed output power of the PV system considering the SB's influence,
e) Updating the SB's SoC from the Kinetic battery model described in (B2) and (B3) in Appendix-B, and
f) Applying the SB's empty or full status (based on $SoC_{SB}(t)$) to update its charging/discharging level, if required.

For each simulation day, the optimization problem in (6) is solved using a binary search algorithm [34] where the SB's capacity is adjusted up and down at each iteration depending on whether the SoC constraints are met. The binary search algorithm ends when the difference between two successive iterations is below a nominated error tolerance level. Fig. A1 illustrates the flowchart of this process.

*2. SBOC Selection*

It was demonstrated in Section V that the results of the optimization studies for both of the MA and RR-based smoothing techniques typically show several days in the year that can be marked as outliers in terms of the SBOC and/or the SIVI. Therefore, from an economic standpoint, there are diminishing investment returns from an SB that covers 100% of all cases (i.e., P100). As such, it is more prudent to select a smoothing level that covers the majority (e.g., 90 or 95%) of the cases (i.e., P90 and P95) and allow curtailment of the PV system for the rest of the time. As such, the SBOC selection is based on the distribution of the single-day SB optimization results using a pre-selected (desired) smoothing level (e.g., P90 or P95).

The steps for selecting the SBOC are as follows:
a) Collecting 1-minute (or higher) resolution solar irradiance data for the site over at least one year,
b) Running the single-day SB optimization for each day in the dataset,
c) Calculating the empirical CDFs and PONE levels for the dataset, and
d) Selecting a smoothing level based on the PONE of the optimization results, and calculating the SBOC according to selected smoothing level.

*B. Modeling of PV and SB Systems*

The output power of the PV system at the coupling point of the inverter at each time step of $t$ ($P_{pv}(t)$) can be calculated by the simplified model of [35]

$$P_{pv}(t) = GHI(t) \times P_{nom} \times k_e \times k_m \times (1 - k_{pt} \times T_{amb}(t)) \times \eta_{inv} \quad (B1)$$

where $k_e$ is the environmental derating factor to account for soiling, dust, etc. [%], $k_m$ is the manufacturer tolerance derating factor [%], $k_{pt}$ is the power-temperature coefficient [%/°C], $T_{amb}(t)$ is the ambient temperature at time step $t$ [°C] and $\eta_{inv}$ is the inverter's efficiency [%]. It is to be noted that this simplified model does not consider the effects of the PV array's tilt angle and azimuth nor the albedo (ground reflectance). The inverter is also assumed to be fully rated to the peak output power of the PV array and no oversupply coefficient is considered (i.e., dc : ac ratio is 1).

The Kinetic battery model developed in [36-38] is used in the simulation model to calculate the amount of energy that can be transferred to and from the SB at each time step. The model reflects the observation that a battery's capacity tends to decrease with an increasing rate of charge or discharge (rate capacity effect). The Kinetic battery model was selected as it is capable of adequately capturing non-linear recovery and rate capacity effects across different battery technologies [19] and thus, has been widely used in similar software tools such as HOMER. The total charge in a battery is divided into an "available" charge ($q_1$ [Ah]), which is energy that is accessible for immediate use, and a "bound" charge ($q_2$ [Ah]), which is energy that is chemically bound, but can be released at a certain rate. The flows of available and bound charges can be described as

$$\frac{dq_1}{dt} = -(I + k_1(1-k_2)q_1 - k_1 k_2 q_2)$$
$$\frac{dq_2}{dt} = k_1(1-k_2)q_1 - k_1 k_2 q_2 \quad (B2)$$

where $I$ is the charge or discharge current [A], $k_1$ is a rate constant at which chemically bound charge becomes available and $k_2$ is the ratio of available charge to total capacity. The differential equations can be solved for each time step $t$ using Laplace transforms as

$$q_1 = q_{1,0} e^{-k_1 \Delta t} + \frac{(q_0 k_1 k_2 - I)(1 - e^{-k_1 \Delta t})}{k_1}$$
$$\quad - \frac{I k_2 (k_1 \Delta t - 1 + e^{-k_1 \Delta t})}{k_1} \quad (B3)$$
$$q_2 = q_{2,0} e^{-k_1 \Delta t} + q_0 (1-k_2)(1 - e^{-k_1 \Delta t})$$
$$\quad - \frac{I(1-k_2)(k_1 \Delta t - 1 + e^{-k_1 \Delta t})}{k_1}$$

where $q_{1,0}$, $q_{2,0}$ and $q_0$ are the available, bound, and total charges at the beginning of the time step and $\Delta t$ is the length of the time step (1 minute = 1/60 hour).

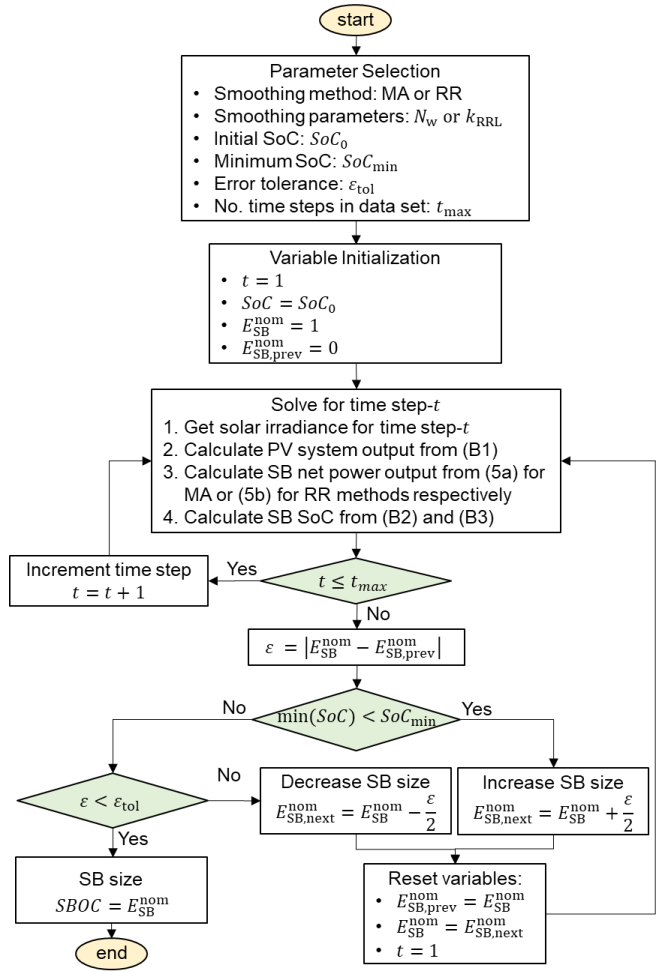

Fig. A1. Flowchart of the single-day SB optimization algorithm.

Table C1. Considered PV parameters in the simulations.

| Parameter | Symbol | Value | Remarks |
|---|---|---|---|
| PV nominal rating | $P_{nom}$ | 1000 [Wp] | |
| Environmental derating factor | $k_e$ | 90 [%] | Assuming light to moderate soiling and dust |
| Manufacturer output tolerance | $k_m$ | 95 [%] | AS/NZS 4509.2 Recommendation [35] |
| Power-temperature coefficient | $k_{pt}$ | 0.38 [%] | Datasheet value for a crystalline silicon module from a Tier-1 manufacturer [41] |
| Inverter efficiency | $\eta_{inv}$ | 95 [%] | Average value for a grid-tied PV inverter from a Tier-1 manufacturer [42] |

Table C2. SB's discharge characteristics [38]

| Discharge time (hours) | 1 | 3 | 5 | 8 | 10 |
|---|---|---|---|---|---|
| Discharge current (A) | 242.4 | 115.7 | 79.8 | 55.23 | 47.01 |

It is to be noted that $k_1$ and $k_2$ are specific to a battery's charging and/or discharging performance, which in turn varies with battery chemistry, capacity and construction. The constants can be estimated using a non-linear least squares algorithm that fits the battery constants with actual charge/discharge performance characteristics, e.g. from a battery manufacturer's datasheet. Also, in the studies of this paper, the flow of charge into and out of the battery is also reduced by the conversion losses from the battery's bi-directional converter. Note that, the kinetic battery model does not consider the effects of temperature, self-discharge and cycle degradation/ageing on the capacity and SoC.

*C. Technical Parameters*

The employed parameters in the modeled PV and battery in Appendix-A are retrieved from [39-42], as given in Table B1 and B2.